\documentclass[conference]{IEEEtran}

\usepackage{cite}
\usepackage{amsmath,amssymb}
\usepackage{amsfonts}
\usepackage{graphicx}
\usepackage{booktabs}
\usepackage[hidelinks]{hyperref}
\usepackage{xcolor}
\usepackage{balance}
\usepackage{tikz}
\usetikzlibrary{arrows.meta, positioning, shapes.geometric, calc, backgrounds, fit}
\usepackage{pgfplots}
\pgfplotsset{compat=1.18}

\newtheorem{defn}{Definition} % definition numbers are dependent on theorem numbers

\begin{document}

\title{Hidden Amplifiers: Cross-Level Risk in Software Supply Chains}

\author{
\IEEEauthorblockN{Rakesh Podder}
\IEEEauthorblockA{
\textit{Colorado State University}\\
Fort Collins, Colorado, USA \\
rakesh.podder@colostate.edu}
\and
\IEEEauthorblockN{Rafael Fabian Gonzalez Arellano}
\IEEEauthorblockA{
\textit{Colorado State University}\\
Fort Collins, Colorado, USA  \\
fabian.gonzalez\_arellano@colostate.edu}
\and
\IEEEauthorblockN{Indrajit Ray}
\IEEEauthorblockA{
\textit{Colorado State University}\\
Fort Collins, Colorado, USA \\
indrajit.ray@colostate.edu}
}
% ------------------------------------------------------------------------------

\maketitle

% ──────────────────────────────────────────────────────────────
\begin{abstract}
Modern software supply chains comprise hundreds of transitive dependencies, yet existing analysis tools operate at either the ecosystem level (dependency graphs) or the code level (static analysis within packages). This separation creates two failure modes. First, false-positive CVE alerts for unreachable code. Second, blind spots for structurally critical micro-dependencies. We introduce \emph{cross-level risk propagation}, a framework that bridges code-level risk metrics with ecosystem-level dependency exposure through a unified risk formula. Preliminary evaluation on 50 packages across npm and PyPI reveals a class of \emph{hidden amplifiers}---micro-dependencies with fewer than 50 methods but over 50,000 dependents---that carry outsized supply-chain risk invisible to all current Software Composition Analysis (SCA) tools. Without cross-level analysis, such packages can harbor exploitable code for years because no current tool considers both internal code structure and ecosystem position simultaneously. These results suggest that cross-level analysis opens a new design space for supply-chain security.
\end{abstract}


% ──────────────────────────────────────────────────────────────
\section{Introduction}
\label{sec:intro}

Software supply-chain attacks have surged in both frequency and impact~\cite{ohm2020backstabber,ladisa2023taxonomy}. The SolarWinds compromise (2020) affected 18,000 organizations through a single build-system intrusion~\cite{peisert2021solarwinds}. Log4Shell (2021) exposed a critical remote code execution vulnerability~\cite{log4shell} in a logging library embedded in millions of Java applications. The \texttt{event-stream} incident (2018) demonstrated that a single malicious maintainer takeover of a micro-dependency can propagate to thousands of downstream projects within days~\cite{garrett2019eventstream}. In response, governments have elevated supply-chain security to a national priority---the U.S. Executive Order 14028 mandates Software Bills of Materials (SBOMs) and supply-chain risk assessment for federal software procurement~\cite{eo14028}.

However, the scale of the problem amplifies these risks. npm alone hosts over 4 million packages, PyPI exceeds 763{,}000, and Maven Central surpasses 600{,}000---collectively serving billions of weekly downloads~\cite{decan2019empirical}. To manage this complexity, SCA tools such as Snyk~\cite{snyk}, Dependabot~\cite{dependabot}, and npm audit have become the state of practice~\cite{rabbi2025understanding,abdalkareem2020impact}. These tools track known vulnerabilities at the \emph{package} level, alerting developers when a dependency has a CVE. However, they treat each package as a black box---they cannot determine whether a vulnerability is \emph{reachable} from the application's code paths~\cite{wang2026reachcheck}, nor can they assess the structural importance of a dependency in the ecosystem.

This black-box limitation creates two concrete failure modes. First, \emph{false-positive overload}. Many dependency-based CVE alerts target vulnerabilities in code that the consuming application never invokes, because SCA tools cannot distinguish reachable from unreachable functions. Developers learn to ignore alerts, burying genuine threats~\cite{kula2018developers}. Second, \emph{hidden amplifier blind spots}. Micro-dependencies such as \texttt{ms} (37 lines of JavaScript, over 829{,}000 dependents) carry enormous structural risk that no code-quality or SCA tool surfaces, because neither analyzes the relationship between internal code structure and ecosystem reach~\cite{wang2026reachcheck,mir2024effectiveness,xu2022insight}. Zimmermann et al.~\cite{zimmermann2019small} documented that individual npm maintainer accounts can impact up to 1 million packages, yet this risk remains invisible to current tooling.

These failures stem from a \emph{level separation}. Ecosystem-level tools do not inspect code, and code-level tools do not consider ecosystem context. Consider \texttt{ms}, a 5-method JavaScript package with 829{,}914 dependents. In 2017, a ReDoS vulnerability was discovered in its core parsing function~\cite{ms_redos}---yet no SCA tool would have flagged \texttt{ms} as high-risk beforehand, because it had no prior advisories and its code is trivially simple. Every dependency update, CVE alert, and transitive dependency decision is a maintenance decision, and current tools do not help developers prioritize structurally important code. A cross-level analysis would have ranked \texttt{ms} among the highest-risk packages based on structural position alone. We introduce \emph{cross-level risk propagation} to bridge this gap. This paper makes three key contributions:
\begin{enumerate}
    \item A \emph{cross-level risk framework} that unifies method-level code metrics with ecosystem-level dependency exposure into a single, comparable score ($\S$~\ref{sec:vision}).
    \item A \emph{multi-layer analysis architecture} that decomposes cross-level risk computation into composable stages, validated by a 16{,}000-LOC open-source prototype for npm and PyPI ($\S$~\ref{sec:framework}).
    \item \emph{Preliminary evaluation results} on 50 packages identifying hidden amplifiers and demonstrating that code-aware ranking substantially improves vulnerability prioritization precision over ecosystem-only approaches ($\S$~\ref{sec:results}).
\end{enumerate}

% ──────────────────────────────────────────────────────────────
\section{Cross-Level Risk Framework}
\label{sec:vision}

\subsection{Core Insight}

The risk posed by a function in a dependency is a product of two independent dimensions. The first is \emph{code-level risk}---how complex, central, and reachable the function is within its own package. The second is \emph{ecosystem exposure}---how many other projects transitively depend on the package. Neither dimension alone is sufficient. Only when both are elevated does a method become a genuine supply-chain concern.

\subsection{Risk Model}

Let $S = \langle P, D \rangle$ be a software supply chain tuple, where $P$ is a set of packages and $D \subseteq P \times P$ is the transitive dependency relation. For each package $p \in P$, let $M(p)$ denote the set of methods extracted via static analysis, and let $G(p) = (M(p), E(p))$ be the intra-package call graph, where $E(p) \subseteq M(p) \times M(p)$ is the set of caller--callee edges. We define two orthogonal risk dimensions and their composition.

\begin{defn}[Code-Level Risk]
Code-level risk $\varphi\colon M(p) \to \mathbb{R}_{\geq 0}$ quantifies a method's structural importance within its package. For each $m \in M(p)$, $\varphi(m)$ is defined as:
\begin{equation}
    \varphi(m) = \underbrace{c(m) \cdot b(m) \cdot r(m)}_{\text{structural}} \cdot\; \underbrace{\tau(m)}_{\text{temporal}}
    \label{eq:composite}
\end{equation}
where $c(m)$ is cyclomatic complexity, $b(m)$ is betweenness centrality in $G(p)$, $r(m)$ is blast radius (transitively reachable callees), and $\tau(m) = \max(\log_2(1 + \mathit{churn}), 1)$ is a temporal factor from git commit frequency.
\end{defn}

\begin{defn}[Ecosystem Exposure]
Ecosystem exposure $\psi\colon P \to \mathbb{R}_{\geq 0}$ quantifies a package's reach across the ecosystem. For each $p \in P$, we instantiate it as $\psi(p) = \mathrm{FanIn}(p)$, the ecosystem-wide transitive reverse-dependency count retrieved from deps.dev~\cite{depsdev}.
\end{defn}

Neither dimension alone is sufficient to characterize supply-chain risk. A complex function in a package with zero dependents poses no supply-chain risk. A trivial getter in a massively-depended-upon package poses minimal code-level risk. We therefore compose them multiplicatively.

\begin{defn}[Cross-Level Risk]
The cross-level risk of method $m \in M(p)$ is:
\begin{equation}
    \mathrm{CLR}(m) = \varphi(m) \cdot f\bigl(\psi(\mathrm{pkg}(m))\bigr)
    \label{eq:clr}
\end{equation}
where $f\colon \mathbb{R}_{\geq 0} \to \mathbb{R}_{\geq 0}$ is a monotone dampening function.
\end{defn}

We choose a multiplicative form because an additive alternative would assign non-zero risk even when one dimension is absent, conflating local code concerns with supply-chain exposure. Fan-in values span six orders of magnitude (1 to 1.19M in our corpus). Without compression, a single high-fan-in package dominates all rankings. We adopt $f(x) = \log_{10}(x + 1)$ as the default, though our sensitivity analysis ($\S$~\ref{sec:results}) shows the ranking is stable across alternatives ($\rho \geq 0.97$). One limitation is that methods with zero betweenness but high complexity may be underweighted. Exploring weighted additive alternatives using ablation study is left for future work.

\subsection{Enabled Analyses}

The cross-level risk model (Definitions~1--3) enables three analyses that no current tool performs.

\textbf{Cross-dependency ranking.} Because CLR (Def.~3) unifies code-level and ecosystem dimensions into a single score, methods from different packages can be directly compared. A method in a low-complexity but massively-depended-upon package can outrank a high-complexity method in a niche package---a comparison impossible with single-level tools.

\textbf{Hidden amplifier detection.} The model reveals an emergent class of high-risk packages invisible to existing tools. We define a \emph{hidden amplifier} as a package $p'$ where $|M(p')| < \theta_M$ and $\psi(p') > \theta_\psi$---few methods but high ecosystem exposure. Such packages are invisible to both SCA tools (which require a known CVE) and code-quality tools (which see only simple code). In our evaluation, $\theta_M = 50$ and $\theta_\psi = 50{,}000$. These thresholds are configurable; the structural gap visible in Figure~\ref{fig:scatter} suggests the hidden amplifier class persists across reasonable threshold choices.

The model also points toward a third capability not yet realized: \emph{reachability-aware CVE filtering}, where alerts are filtered by whether vulnerable functions are reachable from the consumer's call paths. This requires composing call graphs across package boundaries and is scoped to future work ($\S$~\ref{sec:agenda}).

\subsection{Motivating Running Example}

Consider a security engineer reviewing the supply chain of \texttt{express@5.1.0}, a popular Node.js web framework with 45 transitive dependencies. A traditional SCA tool produces CVE alerts for several dependencies, but which alerts demand \textit{immediate attention}?

A code-quality tool can analyze a flagged package like \texttt{qs} in isolation and identify complex parsing functions---but it cannot determine that \texttt{qs} has 57{,}674 dependents, making any exploitable method a supply-chain-wide concern. Conversely, the SCA tool can rank \texttt{qs} by its fan-in, but all methods within \texttt{qs} receive the same score, providing no guidance on \emph{which} functions to audit first.

With cross-level analysis, the engineer can see that \texttt{readStream()} in \texttt{raw-body} scores CLR$=$40.0 (high code complexity $\times$ significant fan-in), while \texttt{compile()} in \texttt{proxy-addr} (CR$=$3.93) ranks nearly as high (CLR$=$21.3) despite lower code-level risk, because \texttt{proxy-addr} has 261{,}831 dependents---$4.5\times$ wider ecosystem reach. The cross-level formula correctly amplifies scores across the ecosystem dimension. Table~\ref{tab:express_clr} summarizes the top-ranked methods. None of these cross-dependency comparisons are possible with existing tools.

\begin{table}[t]
    \centering
    \caption{Top cross-level risk methods in the \texttt{express@5.1.0} supply chain. CR$=$code-level risk (Def.~1), FI$=$fan-in (ecosystem exposure, sourced from deps.dev). %Counts are point-in-time snapshots.
    }
    \label{tab:express_clr}
    \begin{tabular}{llrrr}
        \toprule
        \textbf{Method} & \textbf{Package} & \textbf{CR} & \textbf{FI} & \textbf{CLR} \\
        \midrule
        \texttt{readStream()} & raw-body & 8.33 & 63{,}898 & 40.0 \\
        \texttt{typeis()}     & type-is  & 5.87 & 59{,}995 & 28.0 \\
        \texttt{next()}       & router   & 4.48 & 57{,}674 & 21.3 \\
        \texttt{compile()}    & proxy-addr & 3.93 & 261{,}831 & 21.3 \\
        \texttt{alladdrs()}   & proxy-addr & 2.91 & 261{,}831 & 15.8 \\
        \bottomrule
    \end{tabular}
\end{table}

% ──────────────────────────────────────────────────────────────
\section{Architecture and Implementation}
\label{sec:framework}

\begin{figure}[t]
\centering
\begin{tikzpicture}[
    node distance=0.4cm,
    step/.style={draw, rounded corners, minimum width=1.8cm,
                 minimum height=0.6cm, align=center,
                 font=\scriptsize, text width=2.6cm, inner sep=4pt},
    data/.style={draw, rounded corners, minimum height=0.5cm,
                 align=center, font=\tiny, fill=gray!10},
    arrow/.style={-{Stealth[length=1.5mm]}, semithick},
    byp/.style={-{Stealth[length=1.5mm]}, blue!50, dashed, thin},
    lbl/.style={font=\tiny\itshape, text=black!60},
    band/.style={draw, rounded corners=5pt, dashed, inner sep=6pt},
]

% ── Nodes ─────────────────────────────────────────────────
\node[data] (input) {Root Package};

\node[fill=blue!12, below=0.4cm of input] (resolve)
  {\textbf{Dep.\ Resolution}
   \tiny (registry APIs (npm, PyPI))};

\node[fill=blue!12, below=0.35cm of resolve] (enrich)
  {\textbf{Eco.\ Enrichment}
   \tiny (deps.dev $\cdot$ OSV $\cdot$ Scorecard)};
\node[data, right=0.3cm of enrich] (ext) {\scriptsize deps.dev, OSV};

\node[fill=green!12, below=0.5cm of enrich] (ast)
  {\textbf{AST $\to$ Call Graph $G(p)$}
   \tiny (language-specific walkers)};

\node[fill=green!12, below=0.35cm of ast] (phi)
  {\textbf{Code Risk $\varphi(m)$}
   \tiny (complexity $\cdot$ centrality $\cdot$ churn)};

\node[fill=orange!15, below=0.5cm of phi] (synth)
  {\textbf{CLR $= \varphi(m) \cdot f(\psi(p))$}
   \tiny (cross-package ranking)};

\node[data, below=0.4cm of synth] (output) {Ranked Methods};

% ── Main flow arrows ──────────────────────────────────────
\draw[arrow] (input)   -- (resolve);
\draw[arrow] (resolve) -- (enrich);
\draw[arrow] (ext) -- (enrich);
\draw[arrow] (enrich)  -- (ast);
\draw[arrow] (ast)     -- (phi);
\draw[arrow] (phi)     -- node[right, lbl] {$\varphi(m)$} (synth);
\draw[arrow] (synth)   -- (output);

% ── ψ(p) bypass: enrich → synth (right side) ─────────────
\draw[byp]
  (enrich.east)
  to[out=0, in=0, looseness=2.2]
  node[right, lbl] {$\psi(p)$}
  (synth.east);

% ── Layer bands (background) ──────────────────────────────
\begin{scope}[on background layer]
  \node[band, blue!25,   inner sep=7pt] (b1) [fit=(resolve)(enrich)] {};
  \node[band, green!25,  inner sep=7pt] (b2) [fit=(ast)(phi)]        {};
  \node[band, orange!30, inner sep=7pt] (b3) [fit=(synth)]           {};
\end{scope}

% ── Layer labels (left of each band) ──────────────────────
\node[left=0.15cm of b1, font=\tiny\bfseries, text=blue!70,
      anchor=east] {L1};
\node[left=0.15cm of b2, font=\tiny\bfseries, text=green!60,
      anchor=east] {L2};
\node[left=0.15cm of b3, font=\tiny\bfseries, text=orange!70,
      anchor=east] {L3};

\end{tikzpicture}
\caption{Cross-level analysis workflow. Layer~1 resolves
dependencies and retrieves ecosystem exposure $\psi(p)$.
Layer~2 constructs call graphs and computes code-level
risk $\varphi(m)$. Layer~3 synthesises both dimensions
via Eq.~\ref{eq:clr}. The dashed bypass shows $\psi(p)$
flowing directly from L1 to L3.}
\label{fig:workflow}
\end{figure}
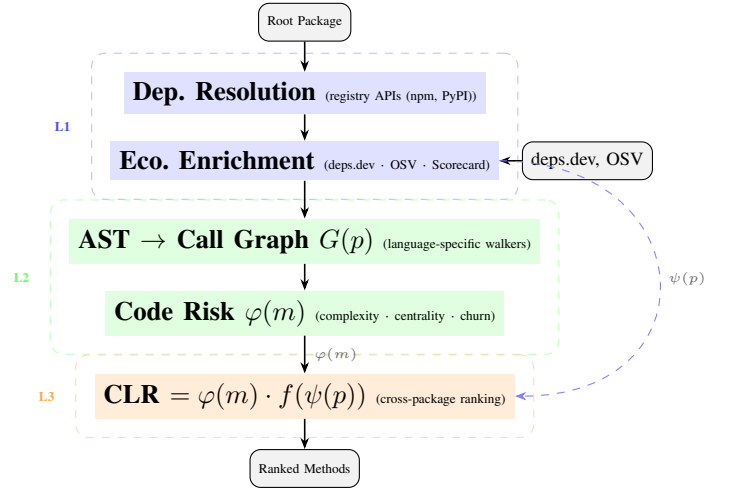

Computing CLR (Def.~3) requires combining data from two independent sources---intra-package call graphs for code-level risk (Def.~1) and ecosystem-wide dependency metadata for ecosystem exposure (Def.~2). However, no existing tool produces both. We decompose the analysis into three composable layers, illustrated in Figure~\ref{fig:workflow}.

\textbf{Layer 1: Ecosystem Analysis} computes ecosystem exposure $\psi(p)$ (Def.~2) for a package $p$. Given a root package, the framework resolves its full transitive dependency tree via registry APIs (npm, PyPI) and enriches each node with fan-in counts from deps.dev~\cite{depsdev}, along with supplementary health indicators---OpenSSF Scorecard~\cite{openssf} scores, libyears freshness~\cite{cox2015measuring}, and known CVEs from the OSV database~\cite{osv}---that inform practitioner triage but are not inputs to the CLR formula. The output is a fully attributed dependency graph.

\textbf{Layer 2: Code Analysis} computes code-level risk $\varphi(m)$ (Def.~1) for each method $m \in M(p)$. For each dependency, the framework clones the source, constructs an AST-based static call graph $G(p)$ using language-specific walkers, and computes per-method risk via Eq.~\ref{eq:composite}. A novel \emph{internal resolution rate} quantifies call-graph completeness.
\begin{equation}
    R = \frac{|\text{resolved internal call sites}|}{|\text{total internal call sites}|}
    \label{eq:resolution}
\end{equation}
where \emph{internal} excludes external and dynamic calls. $R$ serves as a confidence gate. Packages with $R < 0.85$ produce unreliable call graphs, and the framework flags their results accordingly.

\textbf{Layer 3: Cross-Level Synthesis} computes CLR scores (Def.~3) for every method across the dependency tree. Dependencies are first ranked by a \emph{bottleneck score}---defined as $\text{in-degree}(p) \times \text{out-degree}(p)$ within the dependency tree---which identifies packages that mediate the most transitive paths. The top-$N$ bottleneck packages (default $N{=}20$) are selected for method-level analysis, and the resulting per-method risks are unified into a single cross-package ranking.

A prototype implementing this framework (approximately 16{,}000 lines of Python and TypeScript) is publicly available, along with a replication package containing the corpus definition, analysis scripts, and exported datasets.\footnote{Dependency Observatory --- \url{https://github.com/fabgonzalez-dev/oscar-dependency-observatory}, Method Observatory --- \url{https://github.com/fabgonzalez-dev/oscar-method-observatory}, Frontend --- \url{https://github.com/fabgonzalez-dev/oscar-frontend}, Research Data --- \url{https://github.com/fabgonzalez-dev/oscar-research-data}.}

% ──────────────────────────────────────────────────────────────
\section{Evaluation}
\label{sec:results}

We evaluated the prototype on a corpus of 50 packages across npm (35) and PyPI (15). Packages were drawn from the most depended-upon libraries in each registry and stratified into three archetypes. High-impact frameworks (e.g., \texttt{express}, \texttt{flask}), security-critical libraries (e.g., \texttt{axios}, \texttt{urllib3}), and micro-dependencies (e.g., \texttt{ms}, \texttt{inherits}). This curated selection may overrepresent well-maintained packages. A random sample from the full registry would provide a more representative baseline. The corpus includes 10 root packages whose dependency trees contain known CVEs, enabling precision evaluation. The cross-level pipeline produced 825 method-risk entries across all analyzed packages.

For each research question below, we present the motivation, approach, and key findings.

\subsection{\textbf{RQ1: Do hidden amplifiers form a structurally distinct class invisible to current SCA tools?}}

\paragraph{Motivation} Micro-dependencies are invisible to current SCA tools because they have no prior CVEs and their code is trivially simple. If such packages carry outsized structural risk due to their ecosystem position, a new detection mechanism is needed.

\paragraph{Approach} We apply the hidden amplifier thresholds ($|M(p)| < 50$, $\psi(p) > 50{,}000$) to the 50-package corpus and validate with a retrospective analysis of the \texttt{ms} package.

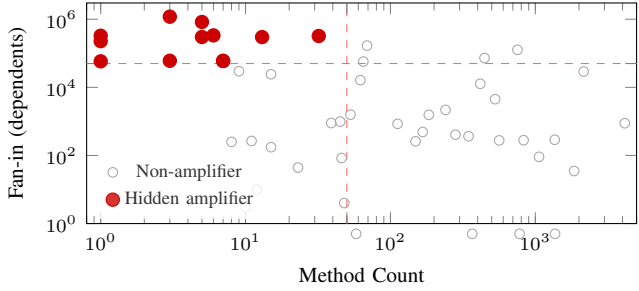
\begin{figure}[t]
    \centering
    \begin{tikzpicture}
    \begin{axis}[
        width=\columnwidth,
        height=4.5cm,
        xlabel={Method Count},
        ylabel={Fan-in (dependents)},
        xlabel style={font=\footnotesize},
        ylabel style={font=\footnotesize},
        xmode=log, ymode=log,
        xmin=0.8, xmax=5000,
        ymin=1, ymax=3000000,
        clip=false,
        xticklabel style={font=\scriptsize},
        yticklabel style={font=\scriptsize},
        legend style={at={(0.02,0.02)}, anchor=south west, font=\scriptsize, draw=none, fill=white, fill opacity=0.8},
        % threshold lines
        extra x ticks={50},
        extra x tick labels={},
        extra x tick style={grid=major, grid style={dashed, red!60}},
        extra y ticks={50000},
        extra y tick labels={},
        extra y tick style={grid=major, grid style={dashed, red!60}},
    ]
    % Non-amplifier packages (gray)
    \addplot[only marks, mark=o, mark size=1.8pt, gray!70] coordinates {
        (167,490) (2155,29123) (45,988) (53,1574) (367,0.5) (779,0.5)
        (1368,0.5) (240,2179) (563,277) (446,72528) (1061,91)
        (1854,35) (528,4490) (62,16282) (58,0.5) (69,167655)
        (65,57024) (149,260) (112,848) (830,280) (184,1558)
        (281,406) (346,367) (4142,882) (753,126582) (417,12742)
        (1361,289) (8,250) (11,266) (23,44) (15,175)
        (15,24264) (48,4) (12,10) (10,3) (46,84)
        (39,894) (9,29762)
    };
    % Hidden amplifiers (red, filled)
    \addplot[only marks, mark=*, mark size=2.5pt, red!80!black] coordinates {
        (3,1187337) (5,829914) (6,334743) (1,327489)
        (32,321932) (5,300803) (13,299023) (1,227359)
        (1,58034) (3,60295) (7,59995) (7,59294)
    };
    \legend{Non-amplifier, Hidden amplifier}
    \end{axis}
    \end{tikzpicture}
    \caption{Method count vs.\ fan-in across 50 corpus packages (log-log). The 12 red points in the bottom-right quadrant ($<$50 methods, $>$50K dependents) are hidden amplifiers---all listed or characterized in Table~\ref{tab:amplifiers}. %The spatial gap between the two populations confirms that hidden amplifiers are a structurally distinct class, not tail outliers.
    }
    \label{fig:scatter}
\end{figure}

\paragraph{Results} Our primary finding is the systematic identification of \emph{hidden amplifiers}---micro-dependencies that score high in cross-level risk despite trivial internal complexity. Of the 50 packages in our corpus, 12 (24\%) qualify as hidden amplifiers ($|M(p)| < 50$ and $\psi(p) > 50{,}000$). This rate reflects our selection of high-fan-in packages. In a random registry sample the prevalence would be much lower. The significant finding is not the rate itself, but that these packages form a \emph{structurally distinct class} visible only through cross-level analysis (Figure~\ref{fig:scatter}). Table~\ref{tab:amplifiers} lists the top examples.

\begin{table}[t]
    \centering
    \caption{Hidden amplifiers: micro-dependencies with $<$50 methods but $>$50,000 dependents. No current SCA tool flags these without a known CVE.}
    \label{tab:amplifiers}
    \begin{tabular}{lrrl}
        \toprule
        \textbf{Package} & \textbf{Methods} & \textbf{Fan-in} & \textbf{LOC} \\
        \midrule
        inherits    & 3  & 1{,}187{,}337 & 23 \\
        ms          & 5  & 829{,}914    & 105 \\
        ee-first    & 6  & 334{,}743    & 94 \\
        escape-html & 1  & 327{,}489    & 46 \\
        depd        & 32 & 321{,}932    & 493 \\
        utils-merge & 1  & 227{,}359    & 8 \\
        fresh       & 3  & 60{,}295     & 89 \\
        \bottomrule
    \end{tabular}
\end{table}

\texttt{inherits}, a 23-line package with 1.19 million dependents and only 3 methods, yields CLR $\approx 12.1$ even with moderate code-level risk. The logarithmic dampening compresses the fan-in advantage: a method in \texttt{raw-body} with $4.2\times$ higher code-level risk scores only $3.3\times$ higher CLR. Figure~\ref{fig:scatter} visualizes this: hidden amplifiers cluster in the bottom-right quadrant, forming a structurally distinct class that no single-dimension tool can identify.

\textbf{Retrospective validation.} The case of \texttt{ms} provides a natural experiment. In 2017, a ReDoS vulnerability was discovered in its core \texttt{parse()} function (fixed in v2.0.0). No SCA tool flagged \texttt{ms} as high-risk before the advisory---it had no prior CVEs and its code is trivially simple. Our framework, applied retroactively, would have ranked \texttt{ms} as a hidden amplifier based solely on structural position.\footnote{deps.dev does not provide historical fan-in snapshots. However, npm contained approximately 350{,}000 packages in early 2017 vs.\ over 4 million today. Proportional scaling yields a conservative 2017 fan-in estimate of $\sim$73{,}000 for \texttt{ms}---well above the 50{,}000 hidden amplifier threshold. With $\varphi \approx 4.0$ for its \texttt{parse()} function, this produces CLR $\approx 4.0 \times \log_{10}(73{,}001) \approx 19.5$, ranking in the top~10 methods across any dependency tree containing \texttt{ms}.} This shows that hidden amplifier detection identifies a class of risk that is invisible to current practice.

\subsection{\textbf{RQ2: Does cross-level scoring produce meaningfully different rankings and improve prioritization?}}

\paragraph{Motivation} If cross-level scoring does not produce meaningfully different rankings from single-dimension approaches, the additional complexity of combining code-level and ecosystem-level analysis offers no practical value.

\paragraph{Approach} We evaluate ranking quality using Precision@$K$ (P@$K$)---the fraction of top-$K$ ranked methods that belong to a dependency with a documented vulnerability---across the 10 root packages with known CVEs. We compare three scoring strategies: CLR (cross-level), CR (code-only), and FI (fan-in only). We present these results as a feasibility demonstration. The sample size precludes statistical generalization.

\begin{table}[t]
    \centering
    \caption{Precision@$K$ across 10 CVE-bearing root packages ($n{=}10$, feasibility only). CLR$=$cross-level, CR$=$code-only, FI$=$fan-in only.}
    \label{tab:precision}
    \begin{tabular}{lccc}
        \toprule
        & \textbf{P@3} & \textbf{P@5} & \textbf{P@10} \\
        \midrule
        CLR (cross-level) & 0.43 & 0.44 & 0.38 \\
        CR (code-only)    & 0.50 & 0.44 & 0.42 \\
        FI (fan-in only)  & 0.20 & 0.20 & 0.23 \\
        \bottomrule
    \end{tabular}
\end{table}

\paragraph{Results} The ranking evaluation reveals two distinct capabilities. First, \textbf{\textit{code awareness improves precision}}: both CLR and CR approximately double ecosystem-only precision across all $K$ values (Table~\ref{tab:precision}). Incorporating code-level metrics improves vulnerability prioritization over fan-in alone. This improvement is shared by CLR and CR because fan-in cannot discriminate between methods within the same dependency.

Second, \textbf{\textit{CLR enables cross-dependency comparison}}---a capability CR lacks entirely. CR can only rank methods \emph{within} a package. CLR produces a single scale \emph{across} packages. CR matches or exceeds CLR on per-root P@$K$ because within a single dependency tree, code-level variation alone is sufficient to discriminate; CR's advantage on this metric does not extend to cross-tree comparison, which CR cannot perform at all. CLR's added value is precisely this cross-project ranking.

The re-ordering is substantial as 24.5\% of methods change rank position when moving from CR to CLR. Ecosystem exposure materially changes prioritization. Evaluating whether this re-ordering correlates with actual vulnerability outcomes requires a larger corpus ($\S$~\ref{sec:agenda}).

%The ranking is also consistent across formulaic choices. 
Testing four dampening functions ($\log_{10}$, $\log_2$, $\sqrt{}$, linear) and computing Spearman's rank correlation coefficient ($\rho$) between the resulting rankings yields $\rho \geq 0.97$ against the default $\log_{10}$. Practitioners can select any variant without sacrificing ranking quality.

\subsection{\textbf{RQ3: How reliable are the data recovery and static analysis components of the pipeline?}}

\paragraph{Motivation} The framework's outputs are only useful if the underlying data recovery and static analysis components produce reliable results. We assess the accuracy of our patch mining technique and the quality of the generated call graphs.

\paragraph{Approach} We evaluate two pipeline components. First, \emph{patch mining}---an automated technique that extracts affected function names from vulnerability fix commit diffs via the GitHub API---is assessed for precision on 45 advisories. Second, call-graph quality is measured via the internal resolution rate $R$ (Eq.~\ref{eq:resolution}) across the 50-package corpus.

\paragraph{Results} For patch mining, 31 of 45 advisories (69\%) yielded function names at 84\% precision. All 15 mined functions found in our call graphs are reachable from the package's public API, confirming these CVEs affect actively used code.

\begin{figure}[t]
    \centering
    \begin{tikzpicture}
    \begin{axis}[
        ybar,
        bar width=16pt,
        width=\columnwidth,
        height=4.0cm,
        ylabel={\# Packages},
        xlabel={Resolution Rate ($R$)},
        ylabel style={font=\footnotesize},
        xlabel style={font=\footnotesize},
        xtick={1, 2, 3, 4, 5},
        xticklabels={$<$0.2, 0.2--0.4, 0.4--0.6, 0.6--0.8, $\geq$0.8},
        xticklabel style={font=\scriptsize},
        yticklabel style={font=\scriptsize},
        ymin=0, ymax=42,
        nodes near coords,
        every node near coord/.append style={font=\scriptsize\bfseries},
        enlarge x limits=0.2,
    ]
    \addplot[fill=blue!50] coordinates {
        (1, 3)
        (2, 3)
        (3, 4)
        (4, 5)
        (5, 35)
    };
    \end{axis}
    \end{tikzpicture}
    \caption{Distribution of internal resolution rate ($R$) across 50 packages. 70\% achieve $R \geq 0.85$. The right-skewed shape reflects a clear divide between statically analyzable and dynamic-dispatch-heavy packages.}
    \label{fig:resolution}
\end{figure}
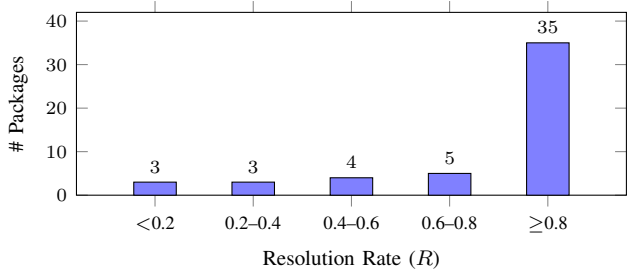

For \emph{call-graph quality}, the internal resolution rate ($R$, Eq.~\ref{eq:resolution}) provides a transparency signal. Figure~\ref{fig:resolution} shows the distribution across our 50-package corpus. The distribution is heavily right-skewed: 70\% of packages achieve $R \geq 0.85$ (suitable for reachability analysis), while a long tail of 20\% fall below $R = 0.60$ (dynamic dispatch). Reporting $R$ alongside risk scores enables practitioners to assess result confidence---epistemic transparency absent from existing tools.

% ──────────────────────────────────────────────────────────────
\section{Discussion}
\label{sec:agenda}
Our results demonstrate feasibility but also reveal limitations and future directions.

\paragraph{Full-graph reachability} Our pipeline exports only the top-$N{=}20$ methods per package. Full call-graph export would enable end-to-end reachability filtering. Early results confirm CVE-affected functions do not cluster in the top-20 (function-level P@5 $= 0.029$). This is an essential next step.

\paragraph{Cross-package call-graph composition} The current reachability scope is \emph{intra}-dependency. Composing call graphs across package boundaries would enable consumer-specific verdicts~\cite{pashchenko2020qualitative}, but requires solving inter-package name resolution---a harder problem due to dynamic imports.

\paragraph{Ecosystem-scale deployment} Our full-clone analysis processes $\sim$3 packages per minute. Scaling to entire registries requires incremental analysis and call-graph caches.

\paragraph{Threats to validity} \emph{Construct:} The multiplicative CLR model assumes risk is zero when either dimension is absent. \emph{Internal:} Our curated corpus overrepresents well-maintained, high-fan-in packages. \emph{External:} Results are limited to npm and PyPI. \emph{Reliability:} Fan-in counts are point-in-time snapshots. $R$ quantifies the static analysis limitation transparently.

% ──────────────────────────────────────────────────────────────
\section{Related Work}
\label{sec:related}

\paragraph{SCA Tools} Production tools (Snyk~\cite{snyk}, Dependabot~\cite{dependabot}, Renovate~\cite{renovate}) operate at the package level, surfacing CVEs via version matching without inspecting code or reachability. \texttt{npm audit} on \texttt{express@5.1.0} produces zero alerts for hidden amplifiers such as \texttt{ms} or \texttt{inherits}. Table~\ref{tab:comparison} shows cross-level analysis against existing approaches.

\begin{table}[!h]
    \centering
    \small
    \caption{Comparison of approaches to supply-chain risk analysis. \checkmark\ indicates capability; --- indicates absence.}
    \label{tab:comparison}
    \begin{tabular}{@{}lcccc@{}}
        \toprule
        \textbf{Approach} & \textbf{Code} & \textbf{Eco.} & \textbf{Reach.} & \textbf{Cross} \\
        \midrule
        SCA tools~\cite{snyk,dependabot,renovate} & --- & \checkmark & --- & --- \\
        Code-quality (e.g., SonarQube\footnotemark) & \checkmark & --- & --- & --- \\
        Eclipse Steady~\cite{plate2015eclipsesteady} & \checkmark & --- & \checkmark$^*$ & --- \\
        PyCG~\cite{salis2021pycg} & \checkmark & --- & --- & --- \\
        Our framework & \checkmark & \checkmark & \checkmark & \checkmark \\
        \bottomrule
        \multicolumn{5}{@{}l@{}}{\scriptsize $^*$Dynamic reachability via bytecode instrumentation.}
    \end{tabular}
\end{table}
\footnotetext{\url{https://www.sonarsource.com/products/sonarqube/}}

\paragraph{Reachability Analysis} Eclipse Steady~\cite{plate2015eclipsesteady} performs dynamic reachability for Java via bytecode instrumentation---the closest prior work. Our framework differs by using static analysis (no test execution required), adding ecosystem-level fan-in scaling, and targeting npm/PyPI. Ponta et al.~\cite{ponta2018beyond,ponta2019projectkb} introduced the ``vulnerable construct'' concept. Our patch mining automates this from fix commit diffs.

\paragraph{Ecosystem Analysis} Decan et al.~\cite{decan2019empirical} compared dependency network evolution across seven ecosystems. Zimmermann et al.~\cite{zimmermann2019small} quantified the blast radius of maintainer compromises in npm. Liu et al.~\cite{liu2022demystifying} traced vulnerability propagation through dependency trees. These works analyze dependency graph structure but do not inspect code within packages---the gap our approach bridges.

\paragraph{Static Analysis and Health Scoring} PyCG~\cite{salis2021pycg} constructs call graphs for Python. Dann et al.~\cite{dann2022modguard} identified challenges for OSS vulnerability scanners. Our prototype adds the resolution rate as a confidence metric absent from prior work. The OpenSSF Scorecard~\cite{openssf} assigns health scores based on maintenance practices but does not inspect code architecture. Our approach evaluates whether code \emph{poses} structural risk rather than whether a project \emph{follows} good practices.

% ──────────────────────────────────────────────────────────────
% Current Limitations section merged into Research Agenda (§\ref{sec:agenda}) to save space.

% ──────────────────────────────────────────────────────────────
\section{Conclusion}
\label{sec:conclusion}

We have introduced cross-level risk propagation, a framework that unifies method-level code metrics with ecosystem-level dependency exposure into comparable risk scores. Evaluation on 50 packages reveals hidden amplifiers invisible to current SCA tools and shows that code-aware ranking substantially improves prioritization precision over ecosystem-only approaches. The retrospective case of \texttt{ms} validates that structural risk signals can provide early warning before disclosure. These results suggest that the current separation between code-quality tools and ecosystem-level SCA is a design limitation, not an inherent boundary---bridging it is both feasible and practically valuable.

% ──────────────────────────────────────────────────────────────
\section*{Acknowledgment}

GitHub Copilot and Claude (Anthropic) assisted with code completion during prototype development ($\S$\ref{sec:framework}) as coding assistance. All research direction, methodology, and interpretation are the intellectual contributions of the human authors.

% GitHub Copilot assisted with code autocompletion during prototype
% development ($\S$\ref{sec:framework}), and Claude (Anthropic)
% assisted with manuscript drafting across all sections and literature
% search. All research direction, methodology, and interpretation
% are the intellectual contributions of the human authors.

% ──────────────────────────────────────────────────────────────
\newpage  % Force references to start on page 6
\balance
\bibliographystyle{IEEEtran}
\bibliography{icsme-2026-visions/ref}

\end{document}